# Characterizing in-text citations in scientific articles: A large-scale analysis

Kevin W. Boyack[1], Nees Jan van Eck[2], Giovanni Colavizza[3], and Ludo Waltman[2]

[1] kboyack@mapofscience.com
SciTech Strategies, Inc., Albuquerque, NM (USA)

[2] [ecknjpvan, waltmanlr] @cwts.leidenuniv.nl
Centre for Science and Technology Studies (CWTS), Leiden University (The Netherlands)

[3] giovanni.colavizza@epfl.ch
Digital Humanities Laboratory, École Polytechnique Fédérale de Lausanne (Switzerland)

## Abstract

We report characteristics of in-text citations in over five million full text articles from two large databases – the PubMed Central Open Access subset and Elsevier journals – as functions of time, textual progression, and scientific field. The purpose of this study is to understand the characteristics of in-text citations in a detailed way prior to pursuing other studies focused on answering more substantive research questions. As such, we have analyzed in-text citations in several ways and report many findings here. Perhaps most significantly, we find that there are large field-level differences that are reflected in position within the text, citation interval (or reference age), and citation counts of references. In general, the fields of *Biomedical and Health Sciences*, *Life and Earth Sciences*, and *Physical Sciences and Engineering* have similar reference distributions, although they vary in their specifics. The two remaining fields, *Mathematics and Computer Science* and *Social Science and Humanities*, have different reference distributions from the other three fields and between themselves. We also show that in all fields the numbers of sentences, references, and in-text mentions per article have increased over time, and that there are field-level and temporal differences in the numbers of in-text mentions per reference. A final finding is that references mentioned only once tend to be much more highly cited than those mentioned multiple times.

## 1 Introduction

The increasing availability of full text from scientific articles in machine readable electronic formats is a development with the potential to greatly impact citation analytics and to significantly improve the accuracy of models of the structure of science. Full text contains information not only on the exact locations of in-text citations within articles, but also on the context in which a citation to previous work is made. Specific problems that can be addressed using full text data include classification of in-text citations by type and function, and improving measures of impact by weighting of citations based on polarity, typology, function, citing location, and perhaps other features as well. Weighting of citations also has the potential to impact our knowledge of the structure of science, in that document clustering (and the resulting maps) could be based on a more accurate measure of the relatedness between documents. These applications, although beyond the scope of this paper, motivate the current work, which studies the characteristics of in-text citations (and associated features) in two large full text databases. A solid understanding of the characteristics of in-text citations is required before advanced applications of full text data, such as those mentioned, can be most fruitfully pursued.

Study of in-text citations and related text from scientific documents using full text sources has a long history. Although both syntactic (the location of references) and semantic (the meaning of references) studies have been pursued (Ding, Liu, Guo, & Cronin, 2013), here we focus primarily on the syntactic aspect. The terminology used in previous studies of in-text citations is not consistent. Thus, to avoid confusion, we define our terminology here. A *reference* is an item in the bibliography or reference list of a document. An *in-text citation* is a *mention* of a reference within the full text of a document. A reference can be mentioned one or more times in a document. Each mention is an in-text citation. We use the terms *in-text citation* and *mention* interchangeably in this article.

Our work examines distributions of in-text citations for two large full text datasets – the PubMed Central (PMC) Open Access subset and a large portion of the Elsevier full text corpus. Using these large and disciplinarily broad datasets, we will show that there are significant variations in the distributions that have not been reported before. We specifically investigate field-level dependencies and report citation count distributions as a function of text progression.

The paper proceeds as follows. We first review relevant literature and then describe our datasets and analysis methods. Results are then reported along with key observations. The paper concludes with a summary, mention of limitations and suggestions for additional work.

## 2 Background

Over the years, many studies of the location or position of in-text citations have sought to identify the relative value of citations as a function of the position or the number of mentions. Implicit among many of these studies is the assumption that references that are more related to the citing article are the more valuable or essential references for that article. In reviewing prior work, we focus on those studies that explicitly include mention location in their analysis.

Early studies were necessarily done by hand with small datasets. In one of the first studies, Voos & Dagaev (1976) examined citations to a set of four highly cited articles, two from biology, one from medicine and one from physics. Despite their very small sample, their findings suggested that a) most mentions come from introduction sections, b) the location and the number of mentions – which early studies often referred to as op. cit. – were both important in determining the value of a citation, c) time was important, and d) different disciplines had different citation patterns. Bonzi (1982) used a set of nearly 500 citations from 31 articles and found that the number of times a work is cited in the text "shows promise of predicting relatedness between citing and cited works".

Cano (1989) sought to study citation function and utility while also examining position. Using 344 references that were coded by function and utility by their authors, they found that references that were mentioned in a perfunctory and negational way were most often peripheral or of low utility. They also found that references classified as organic, conceptual, operational or evolutionary were more typically essential or of higher utility, and that mentions were more concentrated in the first 15% of an article. Hooten (1991) examined 417 citing contexts and found that references with larger numbers of in-text citations seemed more related to the citing paper, and thus more essential, than those with only one in-text citation.

McCain & Turner (1989), using a set of 11 highly cited papers, created an index based on citing location, number of in-text citations, citation utility from citation contexts, and self-citation, finding that papers with a later citation peak (at six years) were more broadly useful than those with an early citation peak. Citations to papers with a later citation peak were more often for methodological advances rather than for experimental results or theoretical concepts. Maričić et al. (1998) examined citation contexts as well as locations using 11% of the mentions to a set of 357 articles, and suggested that references should be valued differently based on the section of the citing article in which they appear. They found references with relatively low "meaning" (or value) to be mentioned predominantly in the introduction, while those mentioned in other sections had higher meaning.

Bornmann & Daniel (2008) examined a set of 350 in-text citations to a set of articles written by grant applicants. Using the IMRaD (introduction, methods, results and discussion) structure, they found that while more mentions appeared in the introduction and discussion sections of citing articles, the methods and results sections were slightly enriched with mentions to articles with higher citation counts. In perhaps the most detailed comparison with ground truth data available, Tang & Safer (2008) surveyed authors of 49 articles in biology and 50 articles in psychology who assessed the mentions in their articles for importance, reason for citation, and relationship to the cited author. They found that reference importance increased proportionally with numbers of mentions and more detailed discussion of the cited document. In addition, the authors considered references mentioned in the methods and results sections to be most important, while those mentioned in the introduction section only were less important than those mentioned in other sections. Hou, Li & Niu (2011) studied 651 biochemistry papers and found that references that shared at least 10 references with the citing paper had, on average, twice as many in-text citations as those with fewer than 10 shared references. In other words, references that were more similar to the citing paper had more in-text citations than those that were less similar.

Several more recent studies have also focused on the value of citations. Wan & Liu (2014) hand-coded 820 references in 40 Association for Computational Linguistics (ACL) papers for citation strength, and found that the average density of citation occurrence (i.e., the smallest distance between neighboring citation occurrences) was the best predictor of strength among the six tested. They applied their regression based on all six features to a much larger set of ACL references and found that only 14% were predicted to be very important. Zhu et al. (2015) surveyed authors of 100 papers (mostly in computer science) who coded 10.2% of references in their papers as "influential". Among the large number of features that were compared with the reference coding, it was determined that the number of in-text mentions was the single feature that best predicted influence. Valenzuela, Ha & Etzioni (2015) annotated 465 references from ACL papers, coding only 14.6% as "important". After training a classifier on their annotated data with a number of features, they found that the number of in-text citations and the number of in-text citations per section were the most predictive features. Jones & Hanney (2016) also found that the percentage of cited articles that are central to the citing article increases with the number of in-text mentions. Zhao, Cappello & Johnston (2017) coded 1473 in-text citations from 14 articles in Journal of the Association for Information Science and Technology (JASIST) by citation function, using three categories that were considered influential and two categories that were considered non-essential. To limit references to those that are influential for the purposes of citation analysis, they suggest that "removing all citation occurrences in the Background and Literature Review sections and uni-citations in the Introduction section appears to provide a good balance between filtration and error rates."

Several recent works have focused once again on the distribution of in-text citations within full text rather than trying to discern the differing value of references. Hu, Chen & Liu (2013) counted the number of mentions per section for 350 articles from Journal of Informetrics, finding that half of the mentions are in the first 30% of the text. They graphically represented distributions of mentions for articles with four, five and six sections. The six-section representation showed the greatest differentiation in the number of mentions per section, with counts decreasing (per kilo-words) through the first four sections, then increasing for the fifth section, and decreasing again for the last section. Ding et al. (2013) also counted mentions by section for 866 JASIST articles, finding that the literature review section had the largest numbers, and that the most highly cited articles were referenced in the introduction and literature review sections.

The largest study of in-text citation distributions, and the most similar to our work, is the recent study by Bertin et al. (2016), who showed distributions of mentions and reference ages as a function of text progression from 45,000 scientific articles published in PLOS journals. Given that the PLOS documents have relatively consistent styles by journal and have well tagged section headers, they were able to characterize these distributions in terms of the IMRaD document style. Specific documents that have sections not in IMRaD order were reordered in IMRaD order, leading to the observation that mentions are most highly concentrated in the introductions of articles, followed by the discussion section. They also showed that reference ages (i.e., time between the publication years of a citing publication and a cited publication) are highest at the beginning of the introduction and in the methods section, decreasing at the end of the introduction and in the results and discussion sections. The fact that these distributions are very similar across PLOS journals led them to characterize these as "invariant" distributions.

Hu et al. (2017) also characterized properties of in-text citations using 350 papers from Journal of Informetrics, and focusing on multiply mentioned references. They found that 25.7% of the references are mentioned more than once with an average of 1.48 mentions per reference, that self-citations are more likely to be multiply mentioned than non-self-citations, and that the number of mentions per reference declines with reference age. Finally, in a related conference paper, we used the two datasets described in the next section to characterize distributions of mentions at the level of the full databases (Boyack, Van Eck, Colavizza, & Waltman, 2017).

In summary, regarding distributions of in-text citations, there is a consensus among previous studies that mentions tend to be more concentrated at the beginnings (e.g., introduction and related work) and endings (e.g., discussion and conclusion) of articles than in the middle sections. There is also a rough consensus that references that are mentioned outside the introductory sections tend to be the most valuable.

**3 Data and Methods**

We have obtained access to two large sources of full text scientific articles that are available in machine readable form – the PubMed Central Open Access Subset (PMCOA) and full text from Elsevier (ELS) journals. Elsevier content is also available in their ScienceDirect product. PMCOA contains roughly 36% of the articles from PubMed Central, and is comprised of a) articles from open access journals and b) articles that are required to be made publicly available under the National Institutes of Health (NIH) public access policy and legislative mandates. Given publisher embargos, inclusion of the latter type of articles is typically delayed 12 months after publication. Nearly all PMCOA articles are also indexed in PubMed. ELS contains articles from almost 3000 Elsevier journals. Given that Elsevier is the largest publisher of journals in the world, ELS is the single largest source of full text scientific articles currently available, and covers most scientific and technical fields, including the social sciences and humanities.

Each full text source was obtained and processed independently, at different times, by different members of our research team (i.e., PMCOA by SciTech Strategies and ELS by CWTS), and using different code, as will be explained below. Despite these differences, the same basic steps were applied to each source: downloading, filtering, parsing, database creation, reference matching, and analysis.

## 3.1 PubMed Central Open Access Subset

The PubMed Central Open Access Subset was downloaded in XML format and processed by SciTech Strategies in October 2015. The subset included data through mid-2015 and contained 1,113,891 individual records, of which 945,279 had an associated PubMed ID (PMID) and at least one reference. Most of the records without PMID were conference abstracts that are not indexed in PubMed. We further limited the data to articles that were classified in PubMed as either a 'journal article' or a 'review', that were published in 1998 or later, and that had at least one reference with a reasonable reference year (defined as being between 1900 and 2015 and where the publication year was no earlier than one year prior to the reference year).

Each XML record was parsed such that individual sections, paragraphs and sentences were identified and numbered. This is important to properly locate each in-text citation. While sections and paragraphs were delimited using XML tags, sentences were delimited and split using the NLTK (http://www.nltk.org/) pre-trained Punkt tokenizer for English. References, along with their bibliographic metadata (including PMID in many cases) were also extracted. In-text citations and their exact positions in the text (in terms of character offsets and text progression centiles within the article) were identified in the sentence level data using reference tags. Multiple in-text citations in the same bracket, whether using author/date or numbered formats, were given the same position in the text, regardless of which reference was listed first. Figure and table captions and footnotes were not considered.

In addition to its centile position within the text, other data were added to each mention including the number of mentions (for the citing article and reference combination), reference age, and the Scopus citation counts to the reference as of the publication year of the citing article. Citation counts were obtained in a two-step process. First, Scopus article IDs were identified for each reference where possible by 1) using the listed PMID for the reference and looking up the corresponding Scopus ID from our matching table, or 2) matching the reference metadata to publication metadata from Scopus. Scopus IDs were identified for nearly 90% of the mentions. Citation counts to each reference were then obtained from our Scopus data tables and added to the data. We note that citation counts are incomplete for references published prior to 1996 since Scopus records only contain references from 1996 onward.

## 3.2 Elsevier Full Text

The Elsevier full text data were obtained in January 2017 and thus included a nearly complete 2016 publication year. Data were downloaded and filtered by CWTS using the following steps. First, the CrossRef REST API was used to identify all publications in Elsevier journals, numbering 8,437,487. Other types of publications, for instance publications in book series or conference proceedings, were not considered. Second, the Elsevier ScienceDirect API (Article Retrieval API) was used to download the identified publications in XML format. Downloading was possible only for the 7,862,859 publications to which Leiden University has access via subscription. For some of these publications, the XML included only metadata rather than full text. Publications without full text were discarded, leaving 6,179,750 XML records. In particular, all publications that appeared before 1998 were discarded, because for almost all of these publications XML formatted full text was not available. Finally, the data were limited to those records that were English-language publications specifically labeled as 'full-length article', 'short communication', or 'review article', leaving a total of 4,821,774 full text records for analysis.

Each XML full text record was parsed to create sections, paragraphs, and sentences. Sections and paragraphs were identified using XML tags. Only major sections were taken into account while subsections were ignored. Sentences could not be identified directly using XML tags. To identify sentences, we used a modified version of the sentence splitting algorithm provided in the BreakIterator class in the Java API. In-text citations and their exact positions in the text (in terms of character offsets and text progression centiles within the article) were identified in the sentence level data using the CWTS parsing algorithms. In-text citations occurring at special locations in the full text of a publication, such as in footnotes and in the captions of tables and figures, were not included in the analysis.

Publications in the ELS database were matched with publications in the Web of Science (WoS) database. Publications were first matched based on DOI. If no DOI-based match was obtained, publications were matched based on the combination of the last name with first initial of the first author, publication year, volume number, and first page number, which together form a relatively unique key for each publication. A match was required for all four fields. Of the publications retained for analysis, 4,503,790 could be matched with a publication in the WoS database. References in the full text publications were also matched with publications in the WoS database. In this case, DOI-based matching could not be applied because references in the ELS database do not include a DOI. Instead, matching was performed based on the four fields mentioned above. We note that the WoS database used by CWTS includes the Science Citation Index Expanded, the Social Sciences Citation Index, and the Arts & Humanities Citation Index. Other WoS citation indices are not included. Publications before 1980 are not included either.

Publications in the ELS database were categorized into five broad fields of science. The fields distinguished in the CWTS Leiden Ranking (www.leidenranking.com) were used for this purpose.

Furthermore, for references in the full text publications, citation counts in the WoS database were determined. This was done only for references to publications indexed in the WoS database used by CWTS.

## 3.3 Descriptive Statistics

Numbers of articles, references and in-text citations for both datasets are given in Table 1, along with other characteristics of the data. Roughly 3% of the references were not included in the PMCOA analysis because they were missing a reference year or did not have a reasonable reference year. In the ELS analysis, all references were included. Although the average article in the ELS dataset is somewhat shorter than the average article in the PMCOA dataset (e.g., in numbers of paragraphs, sentences, characters, etc.), the average numbers of mentions per reference (1.57 for both PMCOA and ELS) and percentage of sentences with mentions (19.9% and 19.8% for PMCOA and ELS, respectively) are very similar. The most substantial difference shown in Table 1 is in the numbers of sections; this difference is explained by the fact that major sections and subsections were counted for PMCOA while only major sections were counted for ELS.

**Table 1. Characteristics of the two full text datasets.**

|  | PMCOA | ELS |
|---|---|---|
| Years covered | 1998-2015 (partial) | 1998-2016 |
| # Publications | 884,557 | 4,821,774 |
| # References | 34,746,187 | 175,156,040 |
| # In-text citations (mentions) | 54,649,985 | 275,337,977 |
| # Mentions w/citation counts | 48,834,690 | 189,482,219 |
| Avg # sections | 12.27 | 5.70 |
| Avg # paragraphs | 39.01 | 32.59 |
| Avg # sentences | 179.19 | 152.45 |
| Avg # sentences w/mentions | 35.67 | 30.14 |
| Avg # characters | 27,582 | 24,588 |
| Avg # references | 39.28 | 36.33 |
| Avg # locations w/mentions | 45.04 | 36.72 |
| Avg # in-text citations | 61.78 | 57.10 |

Full text publications for which a match was obtained between the ELS database and the WoS database were also linked to the five broad fields of science distinguished in the CWTS Leiden Ranking. Publications from 2016 were excluded in this step since we made use of the 2016 edition of the Leiden Ranking, which includes only publications through 2015. The number of publications from 2000 through 2015 that could be linked to one of the five fields is 3,892,083, with a distribution of publications by field as shown in Table 2. *Biomedical and Health Sciences* (BHS) and *Physical Sciences and Engineering* (PSE) are the two largest fields and are roughly the same size, each containing over 35% of the ELS content. *Mathematics and Computer Science* (MCS) and *Social Science and Humanities* (SSH) are the two smallest fields, each with around 6% of the ELS content. Even though the fields differ widely in size, each field is sufficiently large that comparisons between fields should provide robust results.

**Table 2. Number of Elsevier full text papers by Leiden Ranking field (2000-2015).**

| | Abbreviation | # Publications |
|---|---|---|
| Biomedical and Health Sciences | BHS | 1,447,377 |
| Life and Earth Sciences | LES | 568,195 |
| Mathematics and Computer Science | MCS | 237,080 |
| Physical Sciences and Engineering | PSE | 1,430,594 |
| Social Sciences and Humanities | SSH | 208,837 |

## 4 Results

In-text citations (i.e., mentions) from full text articles can be characterized in a number of ways. In this study, we report statistics and distributions related to mentions as a function of time, number of mentions and text progression.

### 4.1 Temporal Trends

Analyses were carried out to identify changes over time in the PMCOA and ELS datasets and in properties related to sentences, references and mentions.

Figure 1 shows that the numbers of documents in PMCOA and ELS have very different temporal trends. ELS contains over 100,000 full text articles published in 1998, which was just over 12% of the number of original research documents (articles and reviews) indexed in WoS that year. ELS full text coverage has grown consistently over the years, stabilizing at roughly 22% of WoS content from 2004-2015, before growing again to 24% of WoS content in 2016. Note that these percentages are based on articles and reviews, and do not include other document types. In contrast, PMCOA contains less than 1% of the articles indexed in PubMed and published prior to 2004, but has rapidly expanded since then to where it contains 21% of PubMed articles published in 2016. PMCOA is highly skewed to recent content, while ELS contains a roughly consistent slice of the scientific literature since 2003. Note that while we had fewer PMCOA articles in 2015 than in 2014 due to the timing of our data acquisition, full year values are shown in Figure 1 (top, red dashed line) for 2015 and 2016 using counts from a more recent query. Figure 1 also shows that the numbers of full text documents in each of five Leiden Ranking fields have increased at roughly similar rates.

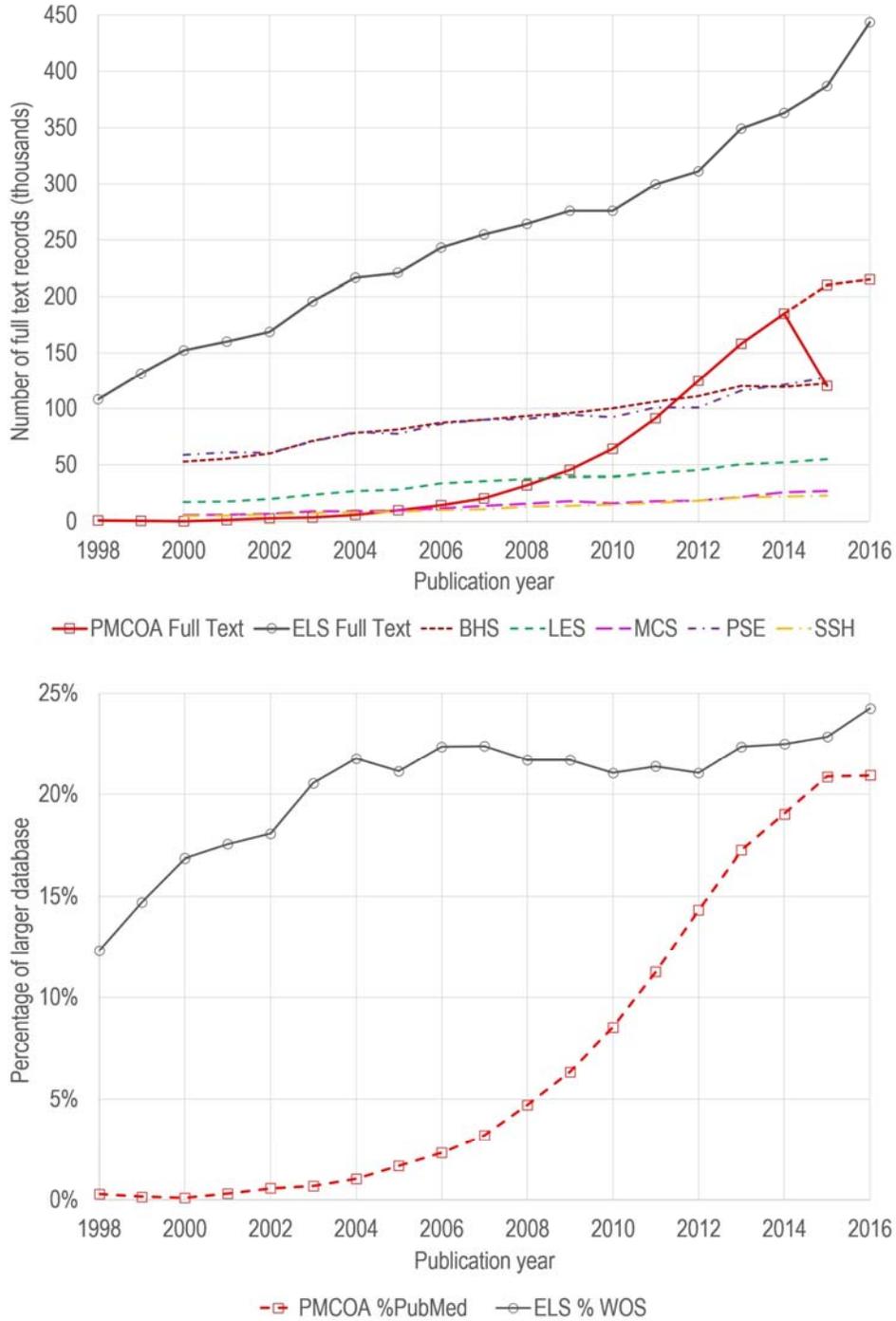

**Figure 1. Full text coverage by year with numbers of documents analyzed by database and field (top) and percentage coverage with reference to a larger database (bottom).**

Table 1 showed that the average numbers of sentences per document were different for the two databases, with 179.2 and 152.5 sentences for the PMCOA and ELS documents, respectively. However, these average numbers do not tell the story. Figure 2 shows that the numbers of sentences vary by field, with SSH documents being longer than those in other fields with over 250 sentences on average in 2015. Documents in the medical (BHS) and physical (PSE) sciences are the shortest with only 150 and 165 sentences, respectively, on average in 2015.

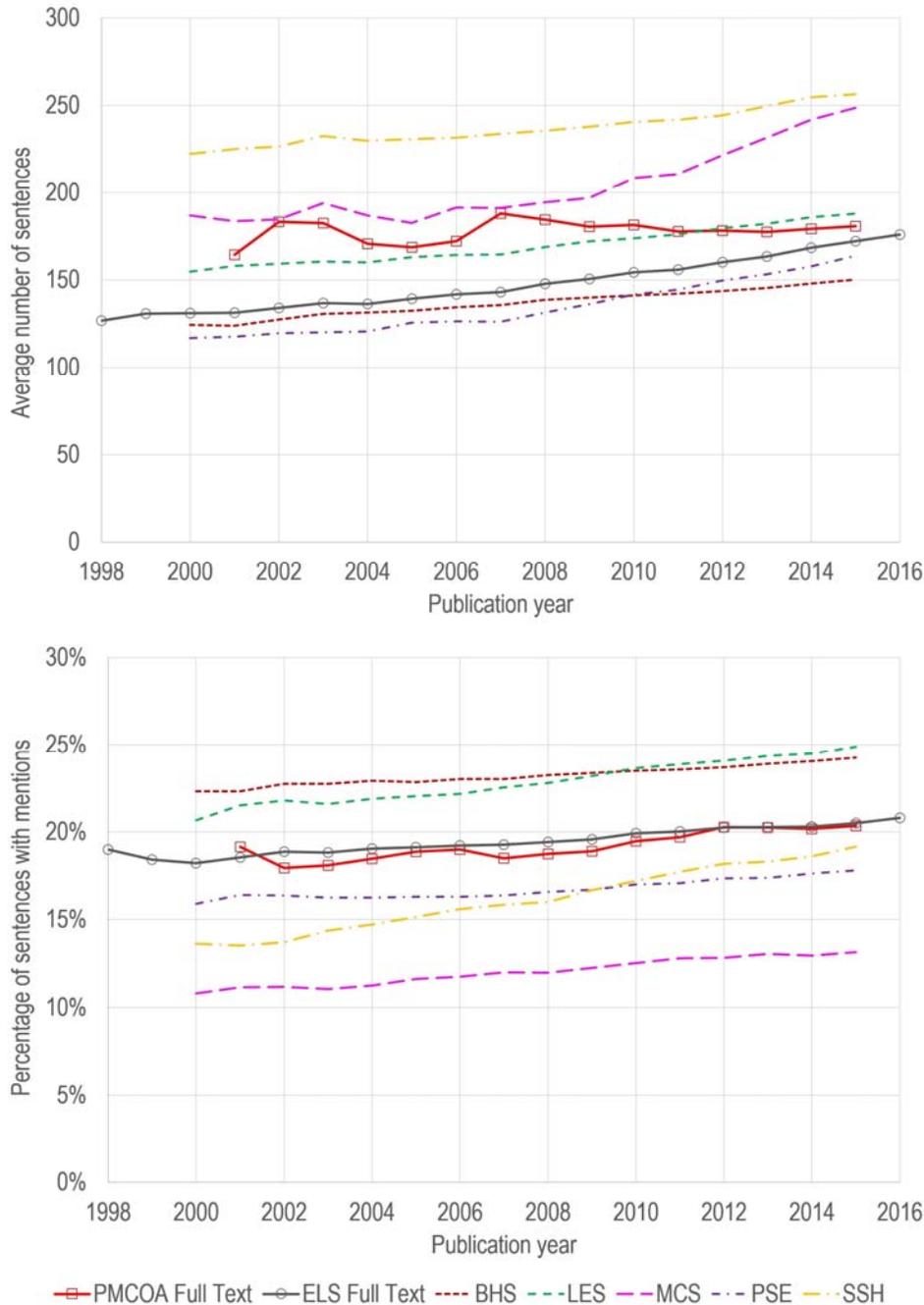

**Figure 2. Average numbers of sentences per document (top) and the percentages of sentences containing mentions (bottom) by year.**

Figure 2 also shows that the average number of sentences per document has been increasing over time for all five Leiden Ranking fields using the ELS data, while the PMCOA data show a roughly constant number of sentences per document. In addition, the percentage of sentences with mentions has been increasing over time for both databases and for all five Leiden Ranking fields. Overall, the increase in numbers of sentences per article (using the larger ELS dataset) may suggest that concerns about salami slicing of publications (Schein & Paladugu, 2001) that were indicated by studies in the 1970s showing a trend toward fewer pages per article (Broad, 1981) are perhaps overstated, and that authors have, for the most part, been packaging their results in "sizable reports" (Bornmann & Daniel, 2007) for the past two decades.

Increases in the number of sentences have been most dramatic for documents in the MCS and PSE fields at around 35% over the sixteen-year period from 2000 to 2015. Increases for BHS and LES have been only 20% over the same time period. It is interesting that even though the numbers of sentences per document has increased over the years, the percentage of sentences with mentions has also increased, indicating a subtle shift in referencing behavior. This may suggest that authors are feeling an increasing need to fully contextualize their research. Increases in the percentage of sentences containing mentions were most pronounced for SSH (39%), and were much lower for other fields. Note that averages and percentages reported in this work were calculated directly at the aggregate level. For instance, in the bottom plot in Figure 2, the percentage of sentences containing mentions was obtained by expressing the number of sentences containing mentions across a set of papers (e.g., the set of all papers in a certain field and year) as a percentage of the total number of sentences across the same set of papers. An alternative approach could have been to calculate percentages at the level of individual papers and to then average these percentages. We did not take this approach.

We had expected the PMCOA and BHS curves to be roughly similar given that the BHS Leiden Ranking field should be similar in scope to what is covered in PubMed. To investigate this further, we identified 11,459 Elsevier documents from 2010-2015 that were in both datasets and found that the PMCOA and ELS versions had on average 170.0 and 189.3 sentences, respectively, per document. These numbers in principle should have been the same for both datasets. The difference must be due to differences in the PMCOA and ELS document formats and/or in the sentence splitting processes used by SciTech Strategies and CWTS. Since this study is focused on syntactic rather than semantic features, we are currently unconcerned about these differences, and will show that while they affect the numbers of sentences, they do not affect analyses based on text progression or numbers of mentions per reference.

Average numbers of references and mentions have both increased over time at a higher rate than that seen for sentences (Figure 3). As it was for sentences, field-level differences exist for references and mentions. LES and SSH have the highest numbers of references and mentions per document, while MCS has the smallest numbers of references and mentions per document. MCS also has the smallest numbers of references and mentions per sentence at just over half the average for all fields. BHS has the largest numbers of references per sentence, while LES has recently overtaken BHS for the largest numbers of mentions per sentence. Growth rates in numbers of references and mentions per sentence have been highest for LES and SSH documents, while growth rates for the other three fields have been relatively modest. The similarity we had expected to see between the PMCOA and BHS curves can be seen in their numbers of references and mentions per document. However, their growth rates are different – numbers for PMCOA documents are growing more slowly than those for ELS documents. This may reflect the selective nature of PMCOA publications, which are either open access publications or those funded by NIH and made available by mandate.

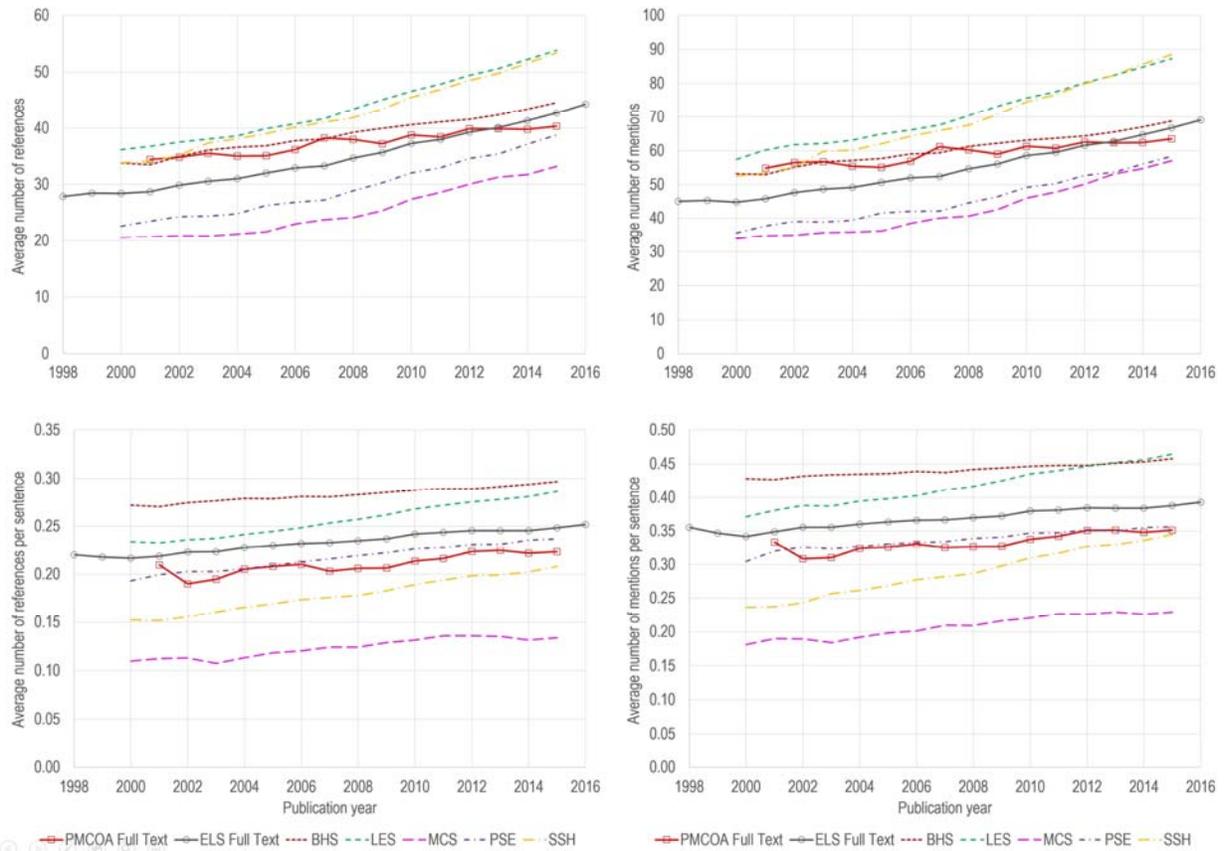

**Figure 3. Average numbers of references (top left), mentions (top right), references per sentence (bottom left) and mentions per sentence (bottom right) by year.**

Despite the growth rates in numbers of references and mentions, the average numbers of mentions per reference are very similar for the two databases and are nearly constant over time, decreasing only slightly from 1.59 in 2001 to 1.57 in 2015 (Figure 4). There are, however, quite significant differences by field. The number of mentions per reference is highest for MCS and increasing slightly, while it is lowest for PSE and decreasing at a substantial rate. Mentions per reference are also increasing at a high rate for SSH. BHS and LES have changed the least over time. This suggests that the degree to which authors engage with the references they cite in their articles has not changed significantly for the health and life sciences (BHS and LES) in the past 15 years, but has changed substantially for the physical, engineering and social sciences (PSE, MCS, SSH). More detailed analysis, perhaps also at a semantic level, will be necessary to understand the reasons for those changes. In addition to the analysis shown here, it was previously shown that the average number of mentions per reference decreases with citation interval (Boyack, et al., 2017).

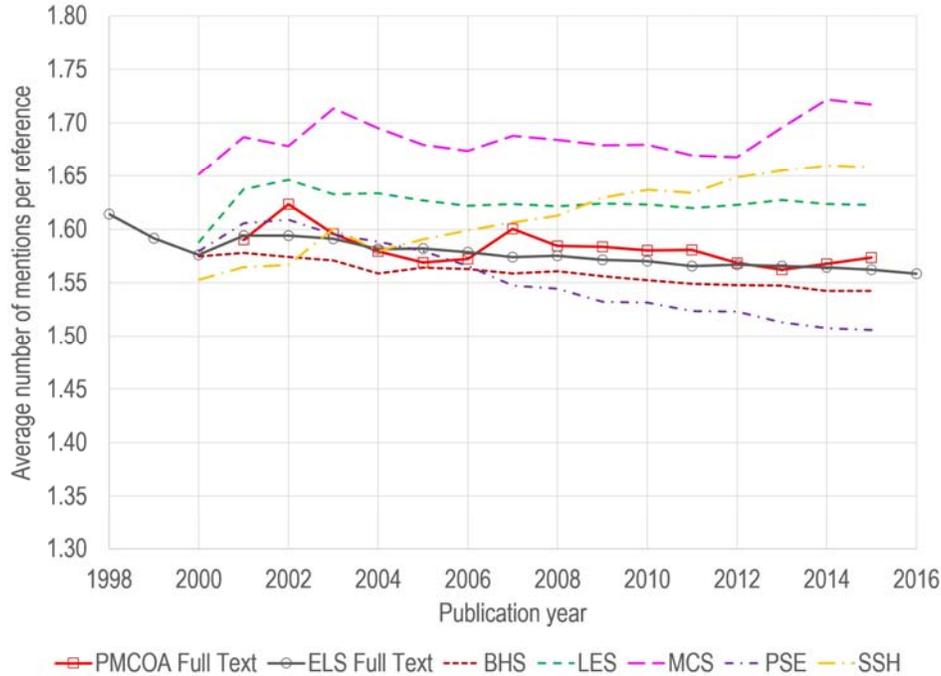

**Figure 4. Average numbers of mentions per reference by year.**

## 4.2 Distributions by Mentions

Figure 2 showed that on average about 20% of sentences contain a reference. Table 3 expands on this and shows the percentage of sentences as a function of the number of mentions and how these vary by field. MCS has the highest percentage of sentences with no references, and the lowest percentage of sentences with mentions for each value of the number of mentions. Less than 2% of sentences in MCS papers have three or more mentions. BHS and LES have the highest percentages of sentences with multiple mentions; yet less than 5% of sentences in these two fields have three or more mentions.

**Table 3. Percentages of sentences as a function of number of mentions contained in a sentence for documents published in 2015.**

| #Mentions | PMCOA | ELS | BHS | LES | MCS | PSE | SSH |
|---|---|---|---|---|---|---|---|
| 0 | 79.66% | 79.48% | 75.75% | 75.12% | 86.84% | 82.20% | 80.83% |
| 1 | 12.56% | 12.72% | 14.63% | 15.22% | 9.00% | 10.92% | 12.37% |
| 2 | 4.75% | 4.01% | 4.98% | 5.04% | 2.24% | 3.37% | 3.68% |
| 3 | 1.54% | 1.73% | 2.19% | 2.20% | 0.85% | 1.47% | 1.54% |
| 4 | 0.70% | 0.84% | 1.05% | 1.04% | 0.41% | 0.75% | 0.71% |
| 5 | 0.32% | 0.44% | 0.54% | 0.54% | 0.22% | 0.42% | 0.36% |
| ≥6 | 0.47% | 0.77% | 0.86% | 0.83% | 0.43% | 0.86% | 0.50% |

It is also useful to know the frequency with which each reference is mentioned. Table 4 shows that 71.5% of references in the PMCOA corpus are mentioned only once, while the number is slightly lower, 69.5%, for the ELS corpus. Given the large sizes of our databases, these numbers are far more definitive than the values of 65.6% and 74.3% reported for JASIST (Zhao, et al., 2017) and Journal of Informetrics (Hu, et al., 2017) documents, respectively. Of the Leiden Ranking fields, PSE has the highest percentage of references mentioned only once, but the numbers across fields do not differ by much. Note also that for the ELS dataset, 1.4% of references were not mentioned in the text at all. Of these, most were mentioned in figure or table captions.

**Table 4. Percentages of references as a function of number of mentions for documents published in 2015.**

| #Mentions | PMCOA | ELS | BHS | LES | MCS | PSE | SSH |
|---|---|---|---|---|---|---|---|
| 0 | | 1.40% | 1.45% | 1.90% | 1.14% | 1.13% | 1.57% |
| 1 | 71.46% | 69.48% | 70.31% | 67.94% | 67.32% | 72.78% | 67.24% |
| 2 | 16.69% | 16.75% | 16.59% | 16.67% | 16.76% | 15.33% | 17.08% |
| 3 | 5.80% | 6.04% | 5.79% | 6.27% | 6.48% | 5.31% | 6.51% |
| 4 | 2.59% | 2.73% | 2.56% | 2.96% | 3.19% | 2.36% | 3.07% |
| 5 | 1.31% | 1.40% | 1.29% | 1.58% | 1.73% | 1.19% | 1.66% |
| 6 | 0.75% | 0.79% | 0.71% | 0.91% | 1.04% | 0.68% | 0.96% |
| 7 | 0.44% | 0.47% | 0.43% | 0.56% | 0.66% | 0.41% | 0.59% |
| 8 | 0.28% | 0.29% | 0.27% | 0.36% | 0.45% | 0.25% | 0.38% |
| 9 | 0.19% | 0.19% | 0.17% | 0.24% | 0.30% | 0.16% | 0.27% |
| 10 | 0.13% | 0.13% | 0.12% | 0.16% | 0.23% | 0.11% | 0.18% |

We also explored citation intervals and numbers of times references had been cited as a function of the number of mentions. Figure 5 shows these statistics for the references and mentions in documents published in 2015. A single year was chosen for this analysis because reference ages and citation counts have both increased over time, and a single year gives a current (rather than an averaged) picture of these distributions. Figure 5 shows that citation interval (i.e., reference age) decreases with the number of times a reference is mentioned. This is true for both databases, and for all fields. However, reference age does vary by field; SSH references are the oldest, while BHS references (which are well mirrored by the PMCOA data) are the youngest at roughly 2.5 years younger than the SSH references.

The statistics also show that references mentioned only once are typically more highly cited than those that are mentioned multiple times. The distributions vary widely by field. For instance, the decrease in citation counts with increasing mentions is quite drastic for the BHS, LES and PSE fields; for these fields citation counts for references mentioned five or more times are less than half that of references mentioned only once. References mentioned only once are also likely to be accompanied by less explanation than those mentioned multiple times (Zhao, et al., 2017). Note also that we have not normalized citation counts with respect to reference age. References mentioned only once are older than references mentioned multiple times, and this may account for part of the difference in citation counts.

We have seen in Figure 5 that references mentioned only once tend to be relatively old and tend to have substantially higher citation counts than references mentioned multiple times. It could be of significant interest to try to explain these observations based on theoretical ideas, either existing ones (e.g., concept symbols, perfunctory citations, etc.) or new ones, about the citation behavior of researchers. We leave this as a topic for future research.

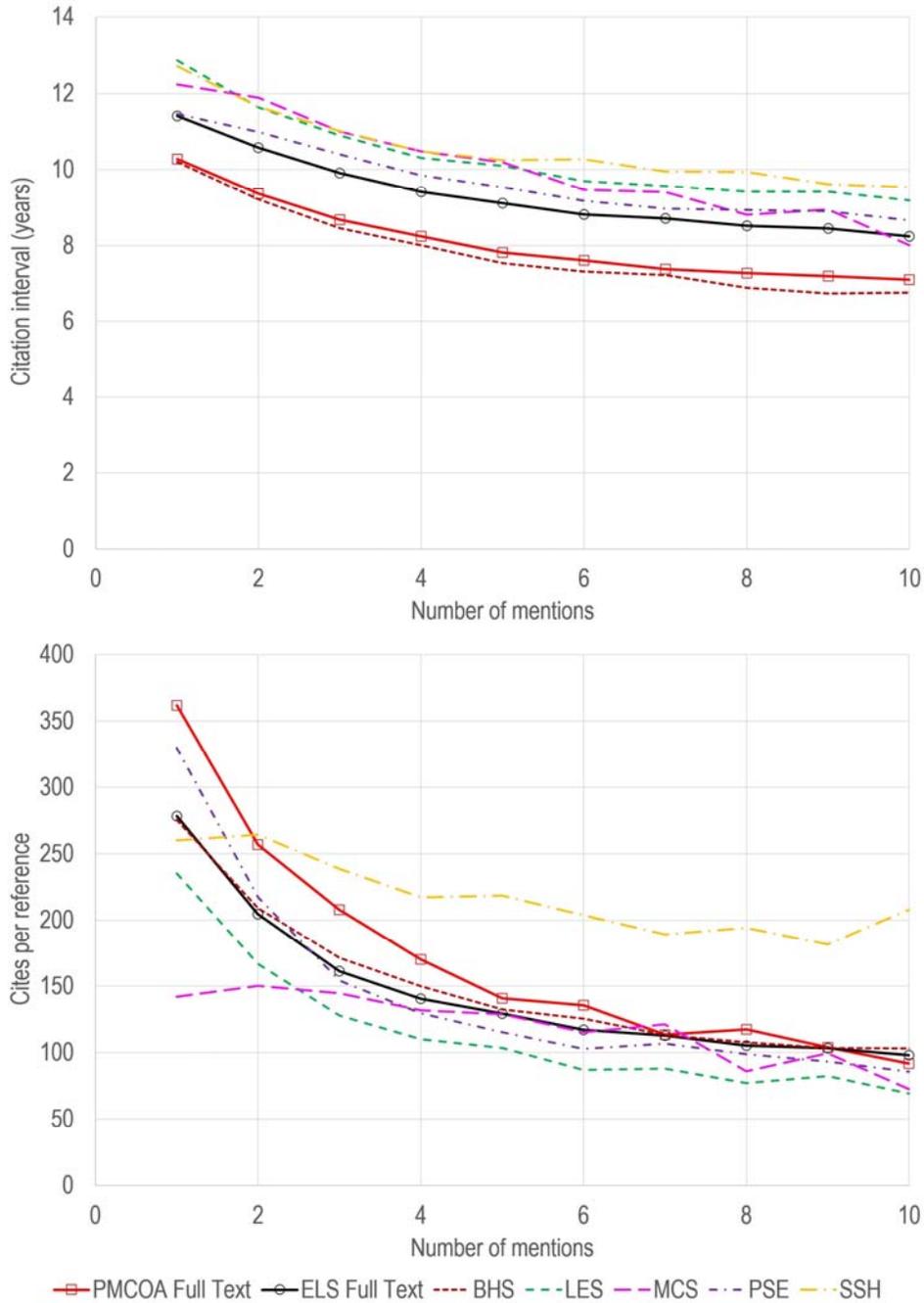

**Figure 5. Citation intervals (top) and citation counts per reference (bottom) as a function of number of mentions for documents published in 2015.**

The MCS and SSH fields show much different behavior. Here, references mentioned twice have slightly higher citation counts than those cited only once, and the decrease with increasing mentions is much more gradual than for the other three fields. Also, it is interesting that SSH references are more highly cited than those in other fields for references mentioned at least twice. That those SSH papers that are cited are on average more highly cited than biomedical papers is perhaps counterintuitive given that lists of the most highly cited papers are typically dominated by biomedicine (Nicholson & Ioannidis, 2012).

One other caveat with respect to citation counts needs to be mentioned. The results in Figure 5 do not include citation counts for references that cannot be identified in WoS (for the ELS data) or Scopus (for the PMCOA data), and might be different if these references and their times cited were known. This is particularly true for SSH because it has a higher fraction of references that cannot be identified than the other fields (Hicks, 2004).

## 4.3 Distributions by Text Progression

While several other studies have aimed to characterize distributions of mentions in terms of the IMRaD structure (Bertin, et al., 2016; Bornmann & Daniel, 2008; Hu, et al., 2013; Maričić, et al., 1998), our study simply characterizes these distributions as a function of text progression. We did not assign mentions to sections due to the lack of uniformity in section naming and ordering across journals. For example, while the ordering implied by IMRaD is likely valid for a significant share of all journals, for other journals it is conventional for the methods section to be at the end of the article rather than after the introduction. In addition, many articles, in particular in SSH journals, have a section structure that differs substantially from IMRaD.

Figure 6 shows the distribution of mentions for the PMCOA and ELS datasets, and for the Leiden Ranking fields, as a function of text progression. Note that, as already mentioned in the discussion of Figure 2, all statistics have been calculated at the aggregate level. The PMCOA and ELS curves both show the same general pattern, with a high level of referencing at the beginning of a document which decreases rapidly to the $25^{th}$ centile where it essentially flattens out, and which is then followed by a secondary peak at around the $80^{th}$ centile before decreasing again at the end of a document. However, this general pattern does not hold for all fields. For instance, documents in the MCS and PSE fields do not have a secondary peak at all, but rather maintain a relatively constant rate of referencing from the $50^{th}$ through $90^{th}$ centiles. Documents in these two fields have the highest fraction of their references (16-18%) within their first five centile portion. In contrast, documents in the SSH field have the lowest fraction of their references (12%) in the first five centiles, and the level of referencing decreases thereafter more gradually than for other fields; the low point in referencing is not reached until the $60^{th}$ centile, after which there is a small secondary peak at the $85^{th}$ centile. The BHS field has a distribution that is similar to that of the PMCOA data, which is what we would expect to see given that they are both medicine-centric.

In a recent article, Bertin et al. (2016) analyzed the distribution of in-text citations of 45,000 documents in PLOS journals. They found that the distribution was similar across all PLOS journals, which led them to characterize these as "invariant" distributions. In our recent conference paper, we showed that the PMCOA distributions are similar to PLOS distributions (Boyack, et al., 2017). However, Figure 6 shows that the PMCOA distribution – and, by extension, the PLOS distribution – is specific to biomedicine, and that there are significant variations in the distributions of references between fields. There is no "invariant" distribution of references in scientific articles. Furthermore, we suspect that there may be significant variations in the distributions within fields that can only be shown with a more granular analysis.

Figure 6 shows the distribution of mentions as a function of text progression for two separate years, 2005 and 2015. Comparison of the curves for the two years suggests that citing behavior in terms of where references are cited in the document text has not changed appreciably over this ten-year period. There is a greater difference between the PMCOA and BHS curves in 2005 than in 2015. We suspect that this difference stems from the relatively small number (10,174) of PMCOA documents in 2005, which were also biased to a particular set of open access journals.

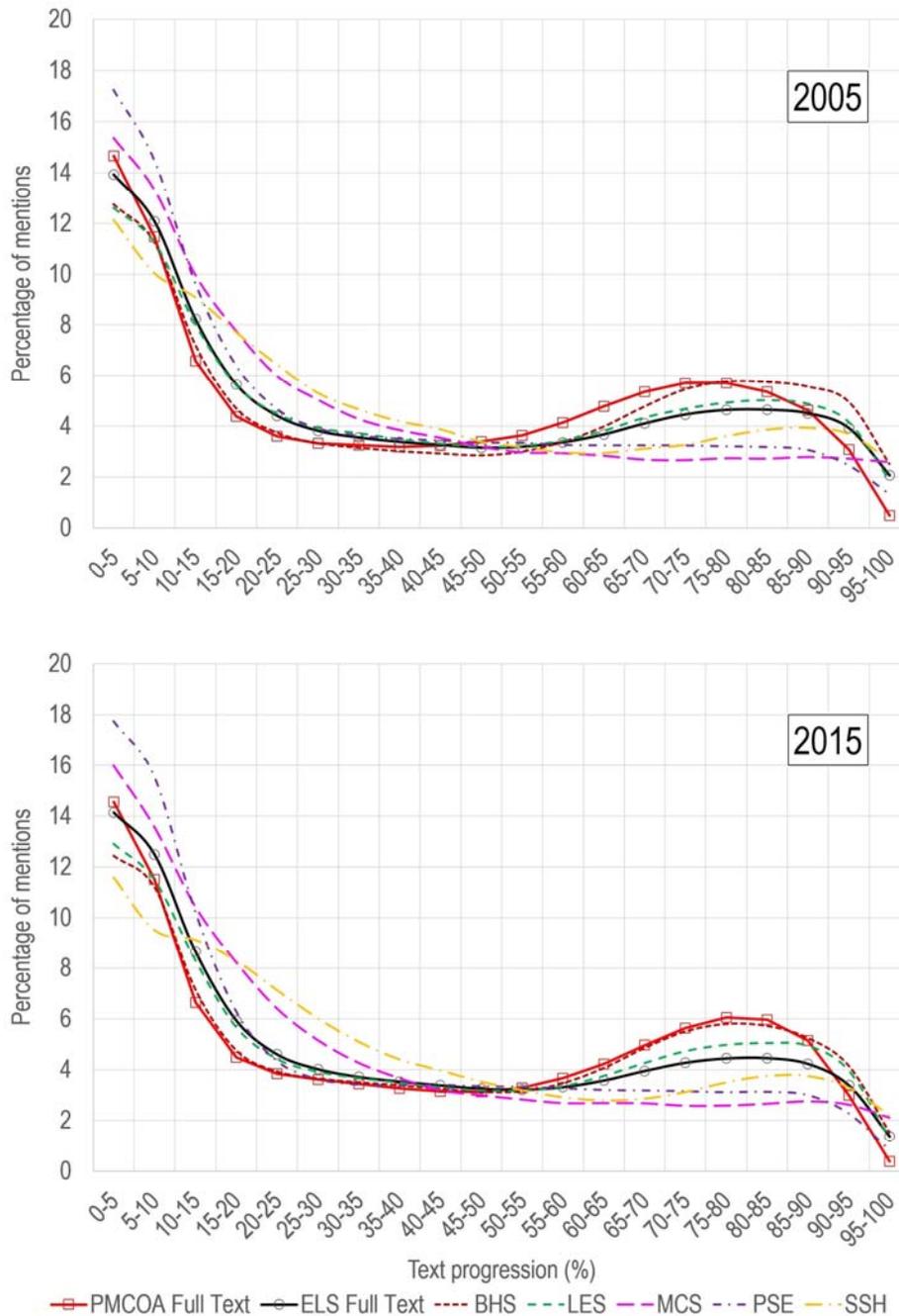

**Figure 6. Percentage of mentions as a function of text progression for documents published in 2005 (top) and 2015 (bottom) using bin widths of five centiles.**

We also investigated citation interval as a function of text progression. Figure 7 shows results for the PMCOA and ELS datasets, and for the Leiden Ranking fields. The PMCOA and ELS curves are similar in nature, with a slight decrease in reference age immediately after the first five-centile bin, followed by an increase to a peak value at the $30^{th}$ to $35^{th}$ centile and then a gentle decay to the end of the article. This pattern is also followed by the BHE, LES, and MCS fields and intuitively makes sense for these empirical fields. The introduction of a paper typically starts by referencing established research, which is followed by references to more recent and directly related research at the end of the introduction. This is followed by methods which are often older, and younger literature is then cited throughout the balance of the paper for comparison purposes. However, the pattern differs for the other two fields. For MCS, the peak does not occur until the $50^{th}$ centile, and for SSH there is no pronounced dip after the first five-centile bin, but rather a gentle and nominal increase to a plateau which extends from the $30^{th}$ to $60^{th}$ centile. This is then followed by the type of decrease seen in the other fields. As we also observed in Figure 5, BHS has the smallest citation interval; references are on average two years older for the other fields. Also, once again, the BHS and PMCOA curves are very similar. Field-level effects are thus seen in citation intervals as well as in the distributions by text progression.

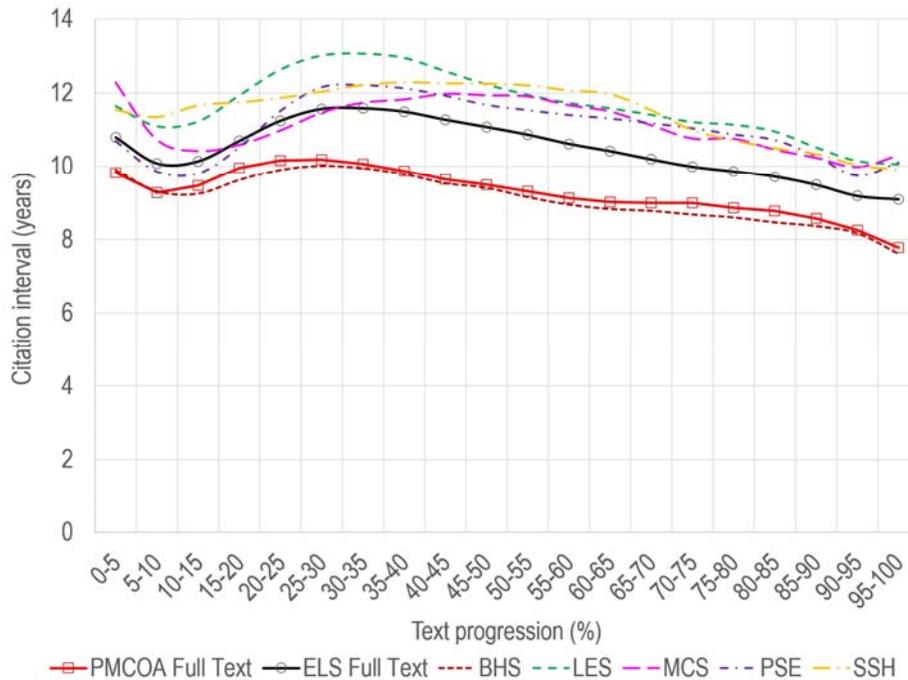

**Figure 7. Citation interval as a function of text progression for documents published in 2015 using bin widths of five centiles.**

Finally, we show the average citation counts for references mentioned as a function of text progression. Figure 8 shows that the PMCOA and ELS curves are similar. References cited at the beginning of an article are more highly cited than those appearing later in the introduction. There then is a huge increase in citation counts to a peak at the $30^{th}$ centile, followed by a decrease, and then another slight increase toward the end of the paper. This final increase is more pronounced for PMCOA than for ELS, probably because PMCOA has a higher share of articles in which the methods section is located at the end of the article.

The citation profiles for the BHS, LES and PSE fields follow this same pattern, although their peak points differ. The peak is reached earlier in PSE articles ($25^{th}$ centile) and later in LES articles ($35^{th}$ centile). Nevertheless, in each of these cases, these peaks still roughly correspond with the traditional location of a methods section, and thus suggest that methods papers are overrepresented among the most highly cited papers in these fields (Van Noorden, Maher, & Nuzzo, 2014). Note that for each of these fields, the peak is roughly three times higher than the valley at the 10th centile.

The other two fields, MCS and SSH, have citation profiles that vary dramatically from those of the BHS, LES and PSE fields. For example, for the SSH field, there is little variation in the citation counts to references over the first 20 centiles, which is followed by a significant increase to a peak at the 60-65th centile at nearly 500 citations per reference. More detailed analysis is required to understand exactly why this curve differs so much from those of the BHS, LES and PSE fields. Nevertheless, it seems clear from these results, and from those in Figure 6, that SSH articles have a different cognitive structure than those in other sciences (Fanelli & Glänzel, 2013). MCS has a profile that is similar to the SSH profile with a peak at the 55th centile, but with far lower citation counts.

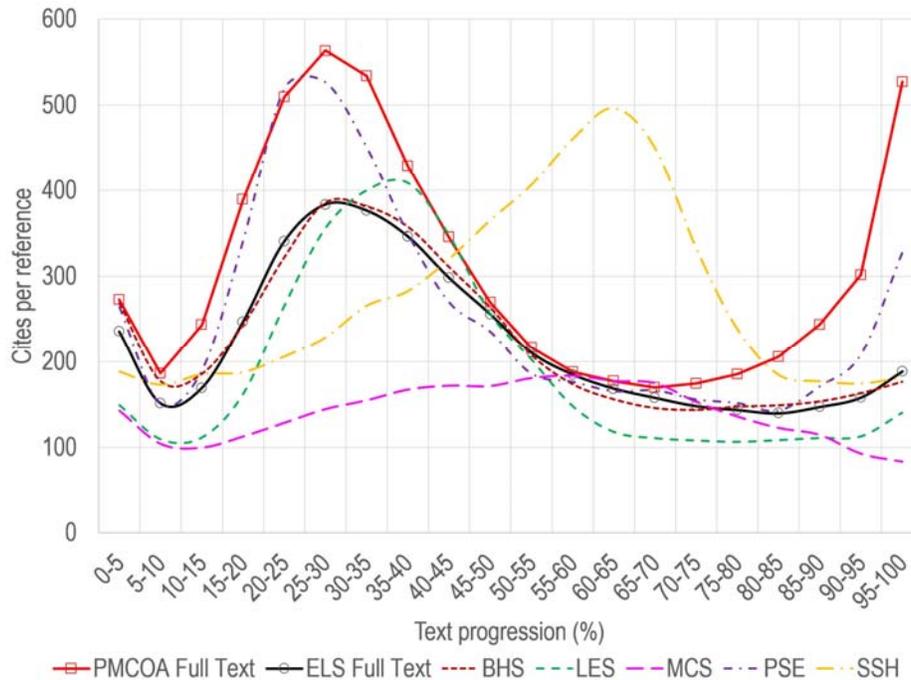

**Figure 8. Citation counts per reference as a function of text progression for documents published in 2015 using bin widths of five centiles.**

One other difference of interest from Figure 8 is that the peak citation counts for references in PMCOA articles are higher than those in BHS articles. While this difference is large at the peak, it is much smaller for references in the first 10 centiles and from the 45th to 80th centiles. Thus, the overall differences are less than what would be suggested by simply comparing peaks. We suspect that an important reason for any differences is that Scopus citation counts are typically higher than WoS citation counts for the same paper since Scopus coverage is broader than WoS coverage. Regarding Figure 8, we note also that the same caveats mentioned in association with Figure 5 apply here as well. These results do not include references that could not be identified in WoS or Scopus, and have not been normalized to account for reference age.

## 5 Conclusions

In this paper, we have analyzed the in-text citations of over five million articles from two large databases – the PubMed Central Open Access Subset and Elsevier journals. Differences between fields of science have been studied using the five fields defined in the CWTS Leiden Ranking. Perhaps most significantly, we find that that there are field-level differences that are reflected in position within the text, citation interval (or reference age), and citation counts of references. These results are fundamentally different from the results of Bertin et al. (2016), who observed that the distribution of references as a function of text progression was relatively constant for PLOS journals. In general, the fields of *Biomedical and Health Sciences* (BHS), *Life and Earth Sciences* (LES), and *Physical Sciences and Engineering* (PSE) have similar reference distributions, although they vary in their specifics. For all three fields, citation counts of references peak at around the $30^{th}$ centile, suggesting that methods papers are more highly cited than other types of papers. This contradicts the finding of Ding et al. (2013) that the most highly cited papers are mentioned in the introduction and literature sections.

The two remaining fields, *Mathematics and Computer Science* (MCS) and *Social Science and Humanities* (SSH), have very different reference distributions from the other three fields. References are more evenly distributed with respect to position in articles in these two fields than in the other three fields, with more references in the $15^{th}$ to $50^{th}$ centile range, and fewer references in the latter half of an article. These fields also differ in that citation counts to references peak at the $60^{th}$ centile rather than at the $30^{th}$ centile. These two observations may suggest that knowledge produced in MCS and SSH relies on previous work in a different way than in the other fields. More detailed work that analyzes semantic as well as syntactic structure will be needed to explore these conjectures.

We have also shown that numbers of sentences, references, and mentions have all increased over time for all fields. The numbers of mentions per reference have remained nearly constant over the past 15 years when considering the entire corpus. However, once again, there are variations by field. Mentions per reference have increased for MCS and SSH, while they have decreased for PSE. Another noteworthy finding is that references mentioned only once are much more highly cited than those mentioned many times. It could be of significant interest to explore possible explanations for this finding.

There are several limitations to this study that should be noted. First, while the study is large, it still covers only a relatively modest share of the articles published in recent years. Additional data from other sources could show different patterns. However, since coverage of the Elsevier full text data is broad from a disciplinary perspective (Boyack, Small, & Klavans, 2013), we do not expect potential differences to be large. Second, only five high level fields were considered, and it is possible that within-field differences may be just as large as the between-field differences shown here. Third, we did not attempt to normalize documents to the IMRaD structure in terms of section names, but simply used textual position as a basis of analysis. Thus, any variations in structural form are not accounted for in our results. We suspect that such normalization would create slightly sharper features in the distributions, similar to what was noted by Bertin et al. (2016) when they unified PLOS documents to the IMRaD structure.

Despite these limitations, we consider the results to be robust. This is by far the largest study of its kind that has been published to date, covering 100 times more articles and references than that of Bertin et al. (2016). Now that distributions have been quantified at a high level, we look forward to pursuing additional studies that explore features of full text articles at more detailed levels, using both syntactic and semantic analyses. Such studies have the potential to influence our understanding of citation theory and behavior, and to have practical influence on applications such as information search and retrieval and accurate modeling of the structure and dynamics of science.

**Acknowledgments**

We thank Mike Patek of SciTech Strategies, Inc. for extraction and fielding of the full text from PubMed Central, and Richard Klavans and Vincent Traag for helpful discussion on our work. Giovanni Colavizza is funded by Swiss National Fund grant number P1ELP2_168489.